# An improvement to the volcano-scan algorithm for atmospheric correction of CRISM and OMEGA spectral data

Patrick C. McGuire[1,2,3], Janice L. Bishop[4], Adrian J. Brown[4], Abigail A. Fraeman[5], Giuseppe A. Marzo[6], M. Frank Morgan[7], Scott L. Murchie[7], John F. Mustard[5], Mario Parente[4], Shannon M. Pelkey[5,8], Ted L. Roush[6], Frank P. Seelos[7], Michael D. Smith[9], Lorenz Wendt[2], Michael J. Wolff[10]

**Abstract**

The observations of Mars by the CRISM and OMEGA hyperspectral imaging spectrometers require correction for photometric, atmospheric and thermal effects prior to the interpretation of possible mineralogical features in the spectra. Here, we report on a simple, yet non-trivial, adaptation to the commonly-used volcano-scan correction technique for atmospheric $CO_2$, which allows for the improved detection of minerals with intrinsic absorption bands at wavelengths between 1.9-2.1 µm. This volcano-scan technique removes the absorption bands of $CO_2$ by ensuring that the Lambert albedo is the same at two wavelengths: 1.890 µm and 2.011 µm, with the first wavelength outside the $CO_2$ gas bands and the second wavelength deep inside the $CO_2$ gas bands. Our adaptation to the volcano-scan technique moves the first wavelength from 1.890 µm to be instead within the gas bands at 1.980 µm, and for CRISM data, our adaptation shifts the second wavelength slightly, to 2.007 µm. We also report on our efforts to account for a slight ~0.001 µm shift in wavelengths due to thermal effects in the CRISM instrument.

---

[1] Formerly at: McDonnell Center for the Space Sciences, Washington University, St. Louis, Missouri, USA.

[2] Department of Planetary Science and Remote Sensing, Freie Universitaet, Malteserstr. 74, Berlin, Germany, D-12249.

[3] Corresp. author: Email: mcguire@epsci.wustl.edu; Phone: (+49)30838-70829; Fax: (+49) 30838-70723.

[4] SETI Institute, Mountain View, CA, USA.

[5] Dept. of Geological Sciences, Brown University, USA.

[6] NASA Ames Research Center, Moffett Field, CA, USA.

[7] Applied Physics Lab, John Hopkins University, Laurel, MD, USA.

[8] Now at: GeoEye Inc., 12076 Grant St. Thornton, CO 80241, USA.

[9] NASA/Goddard Space Flight Center, Greenbelt, MD, USA.

[10] Space Science Institute, Boulder, CO, USA.

A number of scientific results concerning the evolution of the planet Mars have resulted from the analysis of CRISM[1] and OMEGA[2] data, including the mapping by OMEGA of: hydrated sulfates (Gendrin *et al.* 2005), phyllosilicates (Bibring *et al.* 2005) and the polar ice caps (Langevin *et al.* 2005). With its enhanced spatial resolution, CRISM (Murchie *et al.* 2007) has improved the understanding of martian geological evolution by more detailed mapping of sulfates (Murchie *et al.* 2009) and phyllosilicates (Mustard *et al.* 2008; Bishop *et al.* 2008), as well as detections of outcrops of hydrated silica (Milliken *et al.* 2008) and carbonates (Ehlmann *et al.* 2008). Yet despite all of these advances, the CRISM and OMEGA data has not been fully analyzed, in part due to difficulties in correcting for the effects of variability in the observation conditions, caused by variations in atmospheric conditions, surface temperatures and photometric angles. We present here advances to the atmospheric separation algorithm that addresses some of these issues and enables better characterization of the 1.9-2.1 µm wavelength region that contains bands due to $H_2O$ in the surface material.

## 1) Atmospheric correction with the new algorithm

The atmosphere of Mars has a composition that is 95% $CO_2$ (Owen *et al.* 1977), which produces several absorption bands in the near-infrared region (1.0-4.0 µm) (Martin & Barker, 1932). One of these absorption bands is a triad of moderately-deep and narrow bands between 1.9-2.1 µm, which are notable in part because they interfere with the detection of broad features of surface hydration or surface $H_2O$ ice at these same wavelengths. Accurate analysis of these surface features requires removal of the atmospheric absorption features, which can be done by referencing a spectrum of interest to a separate spectrum from a spectrally unremarkable surface where only the atmospheric $CO_2$ feature is present. However, the variable topography on Mars means that the surface pressure of the predominately $CO_2$ atmosphere has a wide range, approximately 1-10 mbars. Furthermore, the atmospheric surface pressure on Mars varies by more than 25% over the course of the martian year, due to the sublimation and eventual recondensation of $CO_2$ in the polar regions (Tillman, Johnson & Guttorp, 1993). These two aspects of spatio-temporal variability of the surface pressure both imply that in order to use the reference spectrum technique to correct for atmospheric absorption by $CO_2$, one must supplement a reference transmission spectrum derived for a specific surface pressure with a technique to adjust it to match the absorption in an observation through a different $CO_2$ column pressure.

---

[1]On-board the Mars Reconnaissance Orbiter spacecraft, the Compact Reconnaissance Imaging Spectrometer for Mars (CRISM) (Murchie *et al.,* 2007) is a visible and near infrared mapping spectrometer operating in the spectral range of 0.362 - 3.92 µm. In this paper, footnotes are used primarily as denoting important scientific or technical details, which can be skipped during a quick reading, but which may be useful to the reader for defining terms or explaining concepts further.

[2]On-board the Mars Express orbiting spacecraft, the Observatoire pour la Minéralogie, l'Eau, les Glaces, et l'Activité (OMEGA) (Bibring *et al.* 2005) is a visible and near infrared mapping spectrometer operating in the spectral range of 0.35 - 5.2 µm.

Indeed, many of the analyses of CRISM and OMEGA data have been undertaken using the so-called 'volcano-scan' technique (Erard & Calvin, 1997; Langevin *et al.* 2005; Murchie *et al.* 2007; Bellucci *et al.* 2007). In the original volcano-scan technique used by the OMEGA and CRISM scientific teams (Langevin *et al.* 2005; Murchie *et al.* 2007; Bellucci *et al.* 2007; which we will refer to as 'version 1' or 'v1'), a reference transmission spectrum is measured by either OMEGA or CRISM by ratioing two spectra acquired nearly simultaneously – one over the summit of Olympus Mons and one from the base of Olympus Mons – on a martian sol when the amounts of ice and dust aerosols were minimal. That ratio gives the transmission of two passes through the atmospheric column between the summit and base of Olympus Mons. This reference is then scaled to a spectrum of interest so that after division by the scaled reference, in the v1 technique, the Lambertian albedo of the surface is forced to be the same at two different wavelengths, one wavelength (1.890 μm) outside the $CO_2$ absorption bands, and one wavelength (2.011 μm) inside the $CO_2$ absorption bands[3]. The following model is used for the atmospheric transmission and surface reflection:

$$I/F(\lambda) = A_L(\lambda)\, cos(\theta_i)\, (T(\lambda))^{\beta} \,. \quad (1)$$

$I/F(\lambda)$ is the radiance, $I(\lambda)$, measured by the CRISM or OMEGA sensor at the wavelength $\lambda$, divided by $F(\lambda) = J(\lambda)/\pi$, where $J(\lambda)$ is the solar irradiance. $A_L(\lambda)$ is the Lambertian albedo at the wavelength $\lambda$, which presumes the Lambertian model for diffuse scattering off of the planetary surface[4], without dependence upon the emission angle, $\theta_e$. The angle of incidence of the sunlight is $\theta_i$. $T(\lambda)$ is the Olympus Mons reference atmospheric transmission spectrum. The exponent, $\beta$, allows for the correction for relative differences in surface pressure between base-to-summit of Olympus Mons to other locations or times on the planet.

---

[3]It is important that the wavelength inside the $CO_2$ bands should be taken at the minimum position of the absorption. The wavelength 2.011 μm is very near the minimum position of the $CO_2$ absorption in this spectral region for the spectral resolutions of the OMEGA and CRISM instruments. The actual minimum position of the $CO_2$ absorption in this spectral region is 2.006 μm. In the case of OMEGA, the minimum is sampled by the spectral channel at 2.011 μm (the OMEGA spectral resolution is about 0.015 μm near this wavelength). In the case of CRISM, the minimum is sampled by the spectral channel at 2.007 micron, which is very close to the minimum of the atmospheric $CO_2$ band.

[4] The assumption here is that the aerosols in the atmosphere are zero. A better nomenclature might be to replace $A_L$ *by* $A_L$* or some other symbol or name in order to remember that aerosols have not been included in the atmospheric correction, but for simplicity, we will sometimes omit the prime, and we will sometimes refer to the measured reflectance as simply Lambert albedo, instead of Lambert albedo*.

The v1 technique of forcing the Lambert albedo to be the same at $\lambda_1$(v1)=1.890 $\mu$m (outside the $CO_2$ absorption bands) and at $\lambda_2$(v1)=2.011 $\mu$m (inside the $CO_2$ absorption bands), assumes that:

$$A_L(\lambda_1(v1)) = A_L(\lambda_2(v1)). \quad (2)$$

For a constant $\theta_i$, this implies that:

$$(I/F(\lambda_1))/(T(\lambda_1)^{\beta}) = (I/F(\lambda_2))/(T(\lambda_2)^{\beta}). \quad (3)$$

The exponent, β, is thus determined from:

$$\beta = ln((I/F(\lambda_1))/(I/F(\lambda_2)))/ln(T(\lambda_1)/T(\lambda_2)). \quad (4)$$

Thus, by determining β on a pixel-by-pixel basis[5], we can correct CRISM or OMEGA spectral data for transmission through the atmosphere, even when there are spatio-temporal variations in surface pressure.

However, the original volcano-scan technique has a number of assumptions which are not universally valid on Mars, including:

1) The spectra of the surface have no differences in albedo between $\lambda_1$(v1) = 1.890 $\mu$m and $\lambda_2$(v1) = 2.011 $\mu$m;
2) The amounts of dust and ice aerosols in the atmosphere over the location of interest are the same as they were in the Olympus Mons region on the sol when the volcano-scan transmission spectrum was measured; and
3) There has been no shift in the wavelength calibration of the instrument (CRISM or OMEGA) since the volcano-scan transmission spectrum was measured.

Assumption #3 will cause difficulties, unless we account for such wavelength changes. For the CRISM instrument, the wavelengths vary by as much as 0.001 µm, due to diurnal and yearly variation of the temperature of the optical assembly of the instrument (Smith *et al.* 2008). The CRISM team is currently improving our processing to account for these wavelength changes, and we discuss this below. In order to address assumption #2, we have been developing and are currently using a more sophisticated atmospheric-correction software pipeline for correcting for spatio-temporal variations in aerosol densities (McGuire *et al.* 2008). This pipeline technique is more computationally intensive than the algorithm discussed here, as it uses historical climatology for the aerosol optical depths and surface pressure, together with a more complete radiative transfer through the atmosphere instead of using transmission spectra. Due to its simplicity and rapid computation, the volcano-scan algorithm is often the preferred approach (over the radiative-transfer calculation) for atmospheric correction.

[5] The v1 volcano-scan algorithm for both CRISM and OMEGA also corrects for a possible spectral slope in the 1.8-2.3 region. This is important because $\lambda_1$ and $\lambda_2$ are so far apart. This formally changes equation 4, though the essence of the resulting calculations is captured by Eq. 4. Equation 4 ensures that the Lambertian albedo at $\lambda_2$ is the same as the (spectral-slope corrected) Lambertian albedo at $\lambda_1$.

Here, we propose a solution to assumption #1 of the volcano-scan atmospheric-correction technique. The surface material in many regions on Mars is composed of hydrated minerals or $H_2O$ ice that naturally produce a broad absorption band depth centered somewhere between 1.9 and 2.1 μm in the surface spectra (e.g. Bibring et al. 2005). The shape of this hydration band is different depending on the material present. For example, phyllosilicates observed in the ancient rock outcrops at Nili Fossae, Mawrth Vallis and elsewhere have a band centered near 1.91-1.93 μm (Mustard *et al.* 2008), while polyhydrated sulfates and monohydrated sulfates from the greater Valles Marineris region exhibit bands in the range 1.93-1.98 and 2.09-2.13 μm, respectively (e.g. Bishop *et al.* 2009; Murchie *et al.* 2009). When present, this hydration band will often produce a difference in albedo between $\lambda_1$(v1) = 1.890 μm and $\lambda_2$(v1) = 2.011 μm. Hence, the original volcano-scan algorithm will introduce significant artifacts. Our proposed solution is to force the spectra to be the same at two new (v2) wavelengths, $\lambda_1$(v2) = 1.980 μm and $\lambda_2$(v2) = 2.007 μm:

$$A_L(\lambda_1(\text{v2})) = A_L(\lambda_2(\text{v2})). \quad (5)$$

This results in the same equation as Eq. 4, but with the version 2 (v2) wavelengths substituted for the version 1 (v1) wavelengths. Figure 1 shows a graphical comparison of the two different volcano-scan techniques for one spectrum that is rich in monohydrated sulfates from Coprates Chasma.[6] See Fueten *et al.* (2009) for a more extensive analysis of this CRISM image in Coprates Chasma from which the spectrum in Fig. 1 was acquired.

The wavelength $\lambda_2$(v2) = 2.007 μm was chosen instead of $\lambda_2$(v1) = 2.011 μm for two reasons: (1) $\lambda_2$(v2) = 2.007 μm is deeper in the $CO_2$ absorption band than 2.011 μm, and hence it will allow more subtle correction for $CO_2$ absorption; and (2) $\lambda_2$(v2) = 2.007 μm is one of the 72 multispectral bands of CRISM whereas 2.011 μm is not. See Murchie *et al.* (2007) for a description of the multispectral mode of CRISM; CRISM "multispectral data" is used for survey observation and, for data volume management reasons, includes a subset of the spectral channels of the full hyperspectral images. Since $\lambda_1$(v2) = 1.980 μm and $\lambda_2$(v2) = 2.007 μm are both multispectral bands for CRISM, this technique should work without adaptation for both multispectral observations and for hyperspectral observations. We have attempted to refine the algorithm further for hyperspectral observations of CRISM, by choosing slightly different wavelengths, but we have found that the wavelengths $\lambda_1$(v2) = 1.980 μm and $\lambda_2$(v2) = 2.007 μm work better than the other possibilities. For OMEGA, since there is not a band at 2.007 μm, we keep the same wavelength for $\lambda_2$: $\lambda_2$(v2,OMEGA)= $\lambda_2$(v1,OMEGA)= 2.011 μm.

---

[6] The previously-mentioned triad of absorption bands in the 1.9-2.1 μm region is evident in the I/F spectrum in Fig. 1A. This 'Fermi' triad of features is caused by a degeneracy of two energy levels in the $CO_2$ molecule (Martin & Barker, 1932). For martian atmospheric conditions, the shape does depend very slightly on the surface pressure and on the temperature of the lower atmosphere. However, the slight pressure-induced or temperature-induced variations of the shape of this triad are very small and have not yet significantly affected the retrievals of the Lambertian albedo spectra in the gas bands in the 1.9-2.1 μm region.

Notably, in the v2 technique, both the wavelengths are inside the $CO_2$ absorption bands between 1.9-2.1 μm. This might appear counter-intuitive at first because one might expect that the technique requires one wavelength outside the gas bands and one wavelength inside the gas bands. However, the measured transmission spectrum at Olympus Mons will automatically account for the two bands both being within the $CO_2$ absorption bands, as long as the band depth of $CO_2$ absorption differs appreciably at the two wavelengths chosen. Furthermore, by choosing this $\lambda_1$(v2) to be 1.980 μm, which is much closer to $\lambda_2$(v2) = 2.007 μm ($\Delta\lambda$ = 0.027 μm), than the wavelengths were before, the restriction that the albedos be the same (Eq.5) is not as stringent as the restriction in Eq.2, where the wavelengths were much further apart ($\Delta\lambda$ = 0.121 μm). Thus, the frequency and/or magnitude of atmospheric-correction artifacts in this spectral region of 1.9-2.1 μm will be much reduced[7]. But most importantly, with the new restriction (Eq. 5), we explicitly allow for differences in albedo between (a) the wavelengths outside the bands of surface hydration at 1.9-2.1 μm and (b) the newly-'restricted' wavelengths of 1.980-2.007 μm, which are often within the bands of surface hydration or surface $H_2O$ ice. This enables quicker and more thorough analyses of surface hydration and surface $H_2O$ ice to be completed.

In Figs. 2-3, we demonstrate the application of this v2 volcano-scan algorithm on several CRISM spectra over sulfate-rich areas of Juventae Chasma and a phyllosilicate-rich area of Nili Fossae, as well as on an OMEGA spectrum in an area east of the Hellas Basin that is rich in $H_2O$ ice (in late southern winter). We show four further examples here in order to demonstrate that this algorithm does indeed function appropriately for different types of spectra at different locations on the planet. More complete studies of the performance of the new algorithm will be forthcoming, as we complete a number of thorough analyses of the performance of the new algorithm for different CRISM and OMEGA images (see Wendt *et al.* (2009), Marzo *et al.* (2009), and Fueten *et al.* (2009) for some examples). In Figs. 2-3, we also compare the results obtained with the v2 algorithm to those obtained with the v1 algorithm. The cases shown in Fig. 2B, Fig. 2C, and Fig. 3 all appear to be much improved. Indeed, for the cases shown in Fig. 2B and Fig. 3, the v2 algorithm eliminates the very strong upward spike at 2.0 μm, which allows for the emergence of a broad and very deep absorption band between 1.9-2.1 μm. Such absorption bands would be expected from either (in Fig. 2B) the hydrated sulfate materials observed in the interior layered deposits in Juventae Chasma (Gendrin *et al.* 2005, Bishop *et al.* 2009) or (in Fig. 3) the presence of $H_2O$ ice remaining from the ice that forms on the surface at the mid-southern latitudes during southern winter (Brown *et al.* 2008). In the latter case, the 1.65 $\mu$m feature is suggestive of larger crystals of $H_2O$ ice, which would indicate that this $H_2O$ ice is a surface feature. Without the 1.65 $\mu$m feature, the presence of absorptions at 1.50 $\mu$m, 2.00 $\mu$m and possibly 1.25 $\mu$m would indicate $H_2O$ ice of smaller grain size, which could be either in the atmosphere as ice aerosols or on the surface as frost.

---

[7] Previous studies with OMEGA and CRISM data often masked out some or all of the wavelengths between 1.95-2.05 μm. Alternatively, in previous studies, spectra were ratioed to 'bland' spectra in the same image that had similar atmospheric artifacts between 1.9-2.1 μm. But this could only be accomplished for a few of the spectra. Other options that have been used for handling the atmospheric artifacts between 1.9-2.1 μm include: (a) applying noise-removal tools to the data cubes, or (b) adding or subtracting a scaled-down version of the transmission spectrum ( $\varepsilon \times T(\lambda)$ ) to the spectrum of interest, where ε is adjusted to smooth out the artifact, as needed.

The case shown in Fig. 2A (CRISM image FRT00009C0A of Mound B) is also improved since if we take the average of those channels between 1.95-2.05 $\mu$m, the v2 algorithm produces an average Lambertian albedo of 0.24. This results in a non-negligible band-depth of ~0.015, or about 6%, accounting for the spectral slope in this 1.8-2.3 $\mu$m region. With the v1 algorithm, it is arguable whether there actually is any absorption between 1.95-2.05 $\mu$m. However, the new results in Fig. 2A, while improved, do indicate that the v2 volcano-scan algorithm does not work perfectly on this image. We expected that if we can better account for the aerosols or for the slight variation of the CRISM wavelengths with time or temperature (assumptions #2-#3 above), then the quality of the atmospheric correction will be even further improved, as discussed in the next section. Furthermore, we are assuming that the underlying spectrum of Lambertian albedo does not change between $\lambda_1$(v2) and $\lambda_2$(v2). There likely is a small spectral slope or a small spectral curvature between these two wavelengths, even if the two wavelengths are relatively close together. Even more advanced versions of the volcano-scan algorithm may be able to correct for spectral slope or spectral curvature. Nonetheless, we feel that our improved handling of assumption #1 of the v1 volcano-scan atmospheric correction algorithm is a result worthy of bringing to the attention of the wider community.

One standard practice used with CRISM and OMEGA data is to ratio two spectra in order to enhance spectral features and estimate the actual band depths in the 1.9-2.1 $\mu$m region. Since the new volcano-scan algorithm eliminates the constraint of a zero difference in albedo between 1.89 $\mu$m and 2.01 $\mu$m,the atmospheric correction more accurately retains the band depths in the 1.9-2.1 $\mu$m region. Therefore, it reduces the need to perform spectral ratios. In fact, the new approach may sometimes result in enhanced atmospheric artifacts between 1.9-2.1 $\mu$m for the spectral ratio, compared to ratios computed with the original approach. This may be caused by the slight variation with time of the CRISM wavelengths, which we address in the next section. However, in most cases, with the v2 volcano-scan algorithm, the spectral ratios are cleaner in the 1.9-2.1 $\mu$m region, as compared to the v1 algorithm, and the underlying spectra more often have more reasonable (non-zero) band-depths in this region.

### 2) Time-dependent volcano-scan transmission spectra

For the CRISM spectrometer, the wavelengths shift during the course of the mission, due to thermal effects on the instrument. We have observed this shift over time while measuring a large number of volcano-scan transmission spectra during the mission. We plot the gas-band portion of 3 of the 16 volcano scan transmission curves in Fig. 4. Note that these volcano-scan spectra effectively record the shifts in the wavelength at the times of the measurement of the transmission spectra. In Fig. 4, the wavelengths shift by < 0.001 μm in the CRISM instrument between the times of the measurement of each of the curves (Smith *et al.* 2008), resulting in transmission curves of slightly different shapes (i.e., slightly different relative band depths for each of the bands)[8,9]. In the future, for the transmission spectrum used in the volcano-scan atmospheric-correction technique for each CRISM image, we will be using the transmission spectrum that best matches each of the thermally-stable periods of the CRISM mission, in terms of the shift of the wavelength.

Preliminary tests of this algorithm are shown in Fig. 5, from an image near an interior layered deposit in Coprates Chasma, and in Fig. 6, for a spectrum from Juventae Chasma and for a spectrum from Nili Fossae. We show two different examples, in order to demonstrate that this technique works in different circumstances. This time-dependent use of volcano-scan transmission spectra (discussed in this section) together with the v2 algorithm (discussed in section 1) are both needed in order to properly correct for atmospheric $CO_2$ in CRISM image FRT00009C0A of mound B of Juventae Chasma; see Bishop *et al.* (2009) and Wendt *et al.* (2009) for a more complete analysis of this image. The pixel location of the spectrum in Fig. 5 is the same as in Fig. 1, and the pixel location of the spectra in Fig. 6 is the same as in Fig. 2.

The spectral comparison in Fig. 5A shows the results with (i) the new time-dependent volcano-scan transmission curve while using the v1 wavelengths, and (ii) the original volcano-scan transmission curve while using the v2 wavelengths. The spectral comparison in Fig. 5B shows the results (both while using the v2 wavelengths) with (iii) the new time-dependent volcano-scan transmission curve, and (iv) the original volcano-scan transmission curve. The intent of Fig. 5A is to demonstrate that the time-dependent shifts of the CRISM wavelengths are not sufficient by themselves to eliminate the artifacts near 2.0 μm.

We make three observations from these preliminary tests. First, the time-dependent transmission curves enabled a smoother Lambert Albedo* curve in the region of 2 μm when processed with the v2 algorithm (see Figs. 5B and 6). Second, when using the time-dependent transmission curves, application of the v1 algorithm still results in artifacts at 2 μm (see Fig. 5A). Third, there are significant differences in the Lambert Albedo* curves shortwards of ~1.45 μm (see Figs. 5 and 6), at the 3-5% level when using the original time-independent transmission curve and the new time-dependent transmission curves, which are caused by the effect of aerosols in the time-dependent transmission curves. We have been developing a technique to account for the effect of these aerosols prior to the standard use of this time-dependent volcano-scan for the atmospheric correction of CRISM data, by empirically removing a small spectral component that scales with wavelength as a power law.

---

[8] It is important to note that a major portion of the temporal variation between the different transmission spectra is not due to the temporal shifts in wavelength, but instead due to temporal changes in the ice and dust aerosol content of the atmosphere near Olympus Mons. The effect of the dust aerosols will tend to be larger at shorter wavelengths. The effect of the ice aerosols will tend to predominate at 1.5 μm and 2.0 μm. We needed to eliminate some of the candidate volcano-scan transmission spectra due to an overabundance of ice or dust aerosols in the region of Olympus Mons.

[9] But even after eliminating the candidate volcano-scan transmission spectra with the largest effects of aerosols, there can still be smaller effects of the ice and dust aerosols on the volcano-scan transmission spectra for which we need to monitor and account.

In the meantime, until this time-dependent volcano-scan technique is further perfected for CRISM data, we will use the original volcano-scan transmission spectrum instead of the set of time-dependent volcano-scan transmission spectra. The allowance for non-zero difference in albedo between 2.01 $\mu$m and 1.89 $\mu$m is a benefit that is too useful not to utilize even if the wavelengths are slightly inaccurate.

### 3) Summary

First, we have proposed and tested a new method for atmospheric separation of OMEGA and CRISM data, adapting the standard volcano-scan technique for more accurate determination of spectral properties of the surface of Mars. Second, we have presented some initial tests of this new algorithm when also accountings for time-dependent shifts in wavelength in the CRISM instrument.

This new method for atmospheric correction allows for a non-zero difference in the Lambertian albedo at $\lambda = 2.007$ $\mu$m relative to $\lambda = 1.890$ $\mu$m, which is a wavelength outside the $CO_2$ absorption bands between 1.9-2.1 $\mu$m. It should reduce the need for the common practice of spectral ratioing, and it should produce more accurate estimates of the depth of bands in the 1.9-2.1 $\mu$m region caused by hydration of the surface or by the presence of $H_2O$ ice on the surface. In other words, there are known surface spectral features of importance that have non-zero albedo difference between the original two wavelengths with their rather large separation ($\Delta\lambda$(v1) = 0.12 $\mu$m). Additionally, minerals that have non-zero albedo difference between the v2 wavelengths (with $\Delta\lambda$(v2) = 0.027 $\mu$m) are not nearly as common, at least in non-polar regions. Due to the similarity of the spectra of $CO_2$ ice and $CO_2$ gas, this algorithm still will suffer when there is $CO_2$ ice on the surface. Furthermore, when there are significant aerosols, the new (v2) algorithm still presents imperfect results. Nonetheless, we expect the application of this new technique to allow for quicker and more thorough analysis of OMEGA and CRISM data.

**Acknowledgements:** PCM acknowledges support from and conversations with Raymond Arvidson, Gerhard Neukum, Selby Cull, Sandra Wiseman, Kim Lichtenberg, Bethany Ehlmann, Ernst Hauber, Tom Stein, Lars Arvidson, and Margo Mueller. The work by PCM has been funded by a Robert M. Walker senior research fellowship from the McDonnell Center for the Space Sciences and by a Humboldt Research Fellowship. The authors from institutions in the USA acknowledge support from NASA funds through the Applied Physics Laboratory, under subcontract from the Jet Propulsion Laboratory through JPL Contract #1277793. LW has been supported by the German Space Agency (DLR) within the Mars Express project.

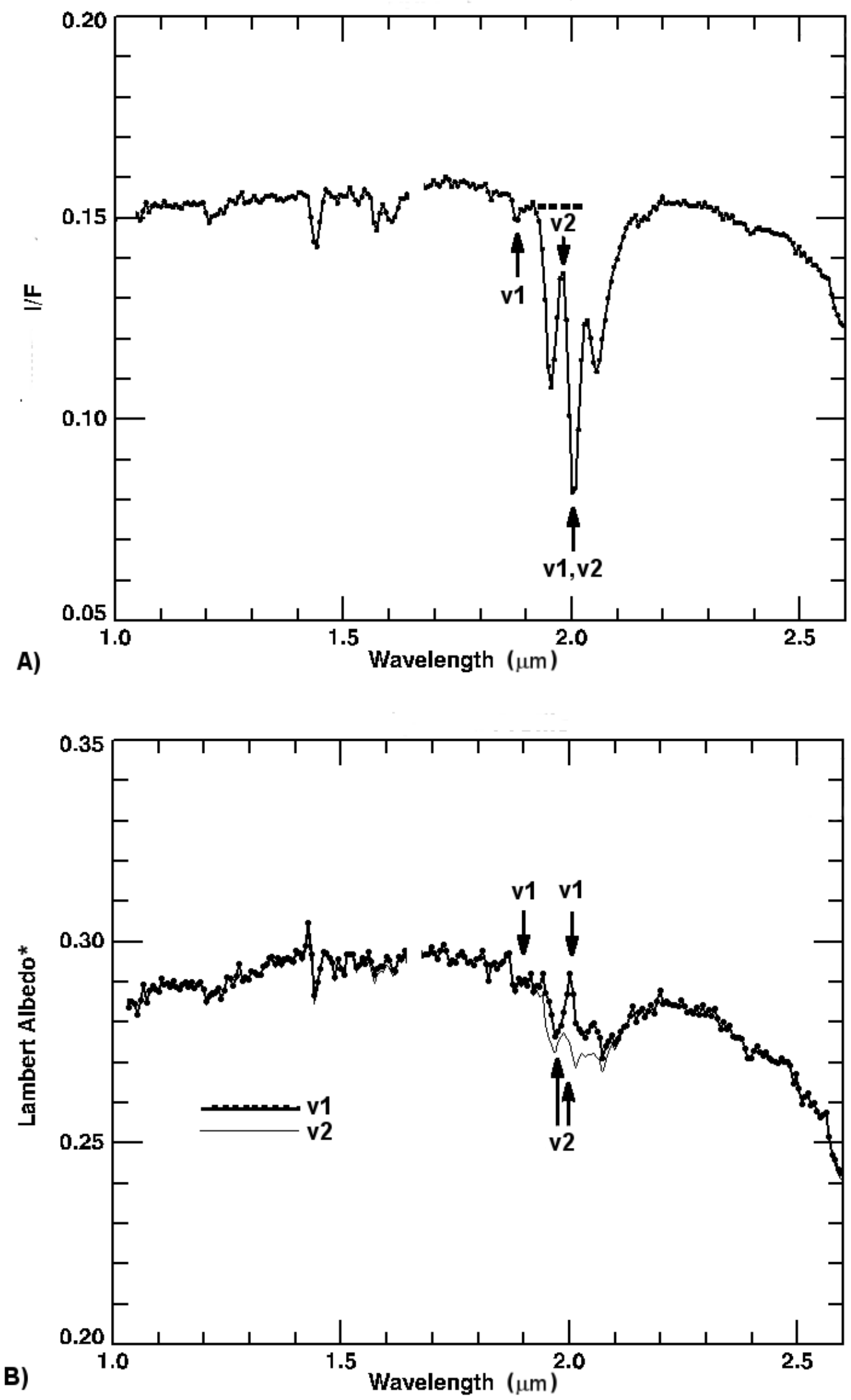

**Figure 1)** Comparison of the original (v1) and new (v2) volcano-scan techniques applied to an I/F spectra. Panel A is from a pixel in CRISM image HRL00003752 of Coprates Chasma. The three wavelengths of interest, $\lambda_1$(v1)=1.890 μm, $\lambda_1$(v2)=1.980 μm, and $\lambda_2$(v2)=2.007 μm, are indicated by arrows. The original $\lambda_2$(v1) = 2.011 μm is not indicated since it is so close to $\lambda_2$(v2). The dashed line indicates the flattening that the original volcano-scan algorithm would provide. In Panel (B), the solid gray line shows the result of the v2 volcano-scan algorithm, together with the two wavelengths of interest. The black line shows the result of the v1 volcano-scan algorithm, together with the two wavelengths of interest. The vertical range of all panels is the same (0.15), for ease of comparison, even though the minimal and maximal values are different. Note that the atmospheric artifacts between 1.9-2.1 μm with the v2 algorithm are reduced in amplitude by factor of 4-5 compared to the v1 algorithm. Also note, for both versions, the correction of smaller $CO_2$ features between 1.555-1.665 μm, and a mis-correction of the $CO_2$ feature at 1.45 μm.

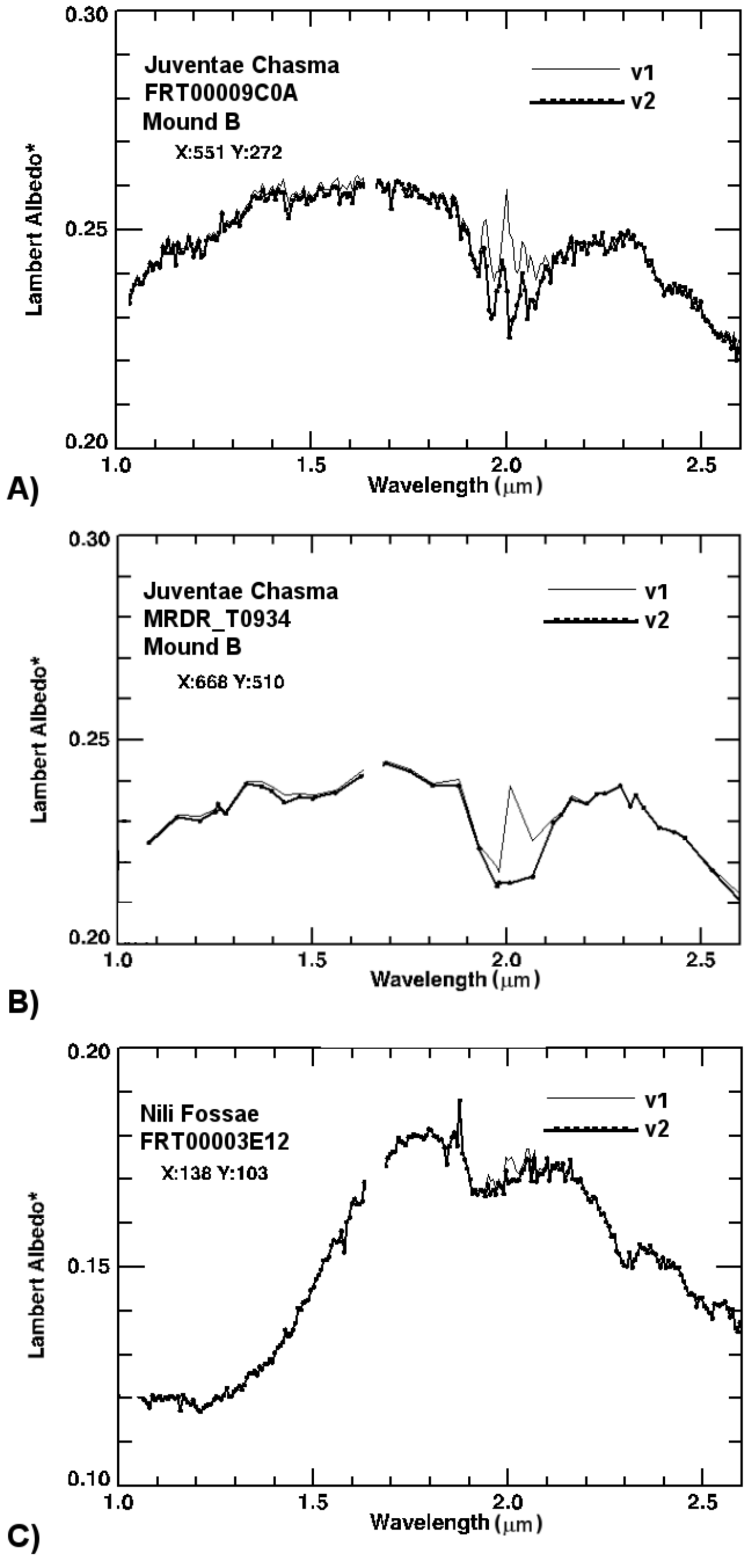

**Figure 2)** A comparison of corrected spectra using the v1 and v2 volcano-scan algorithms is shown for portions of three different CRISM spectra for hyperspectral and multispectral observations of sulfate-rich areas of Juventae Chasma, and a hyperspectral observation of a phyllosilicate-rich area of Nili Fossae. All should show mineralogical absorption in our wavelength range of interest (1.9-2.1 μm).

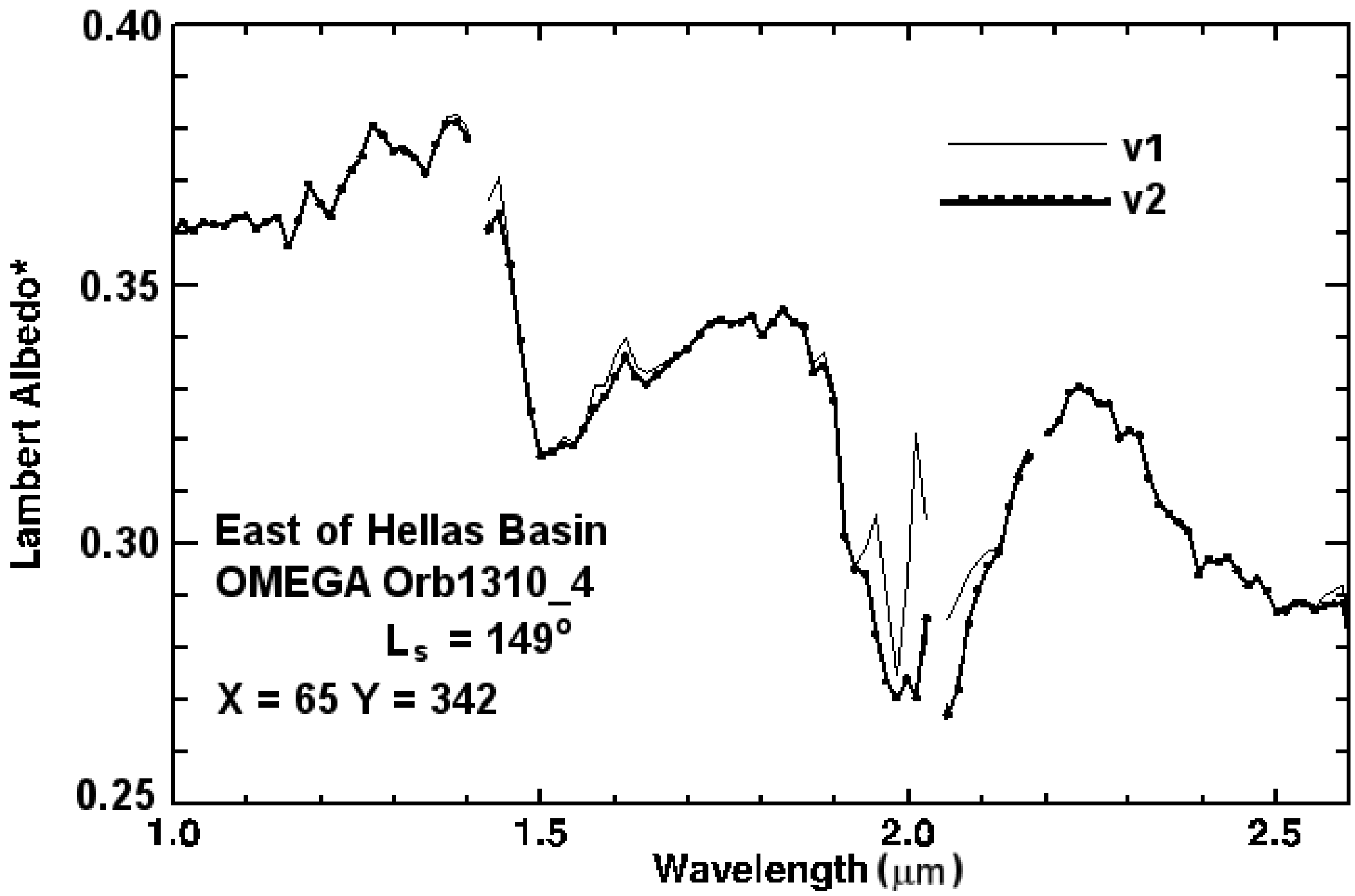

**Figure 3)** A comparison of the v1 and v2 volcano-scan atmospheric-correction algorithms, when applied to a spectrum, primarily due to $H_2O$ ice, acquired by OMEGA in terrain east of Hellas Basin. In the region from 1.9-2.1 μm, with the exception of the channel at 2.02 μm, the artifacts due to atmospheric correction are reduced by a factor of 10-50 when using the v2 algorithm as compared with the v1 algorithm. For $H_2O$ ice, there should be a single broad absorption band centered at 2.0 μm. We speculate that the channel at 2.02 μm has some fixed pattern noise in this image. For clarity, bad channels at 1.43 μm, 2.03 μm and 2.18 μm are not shown.

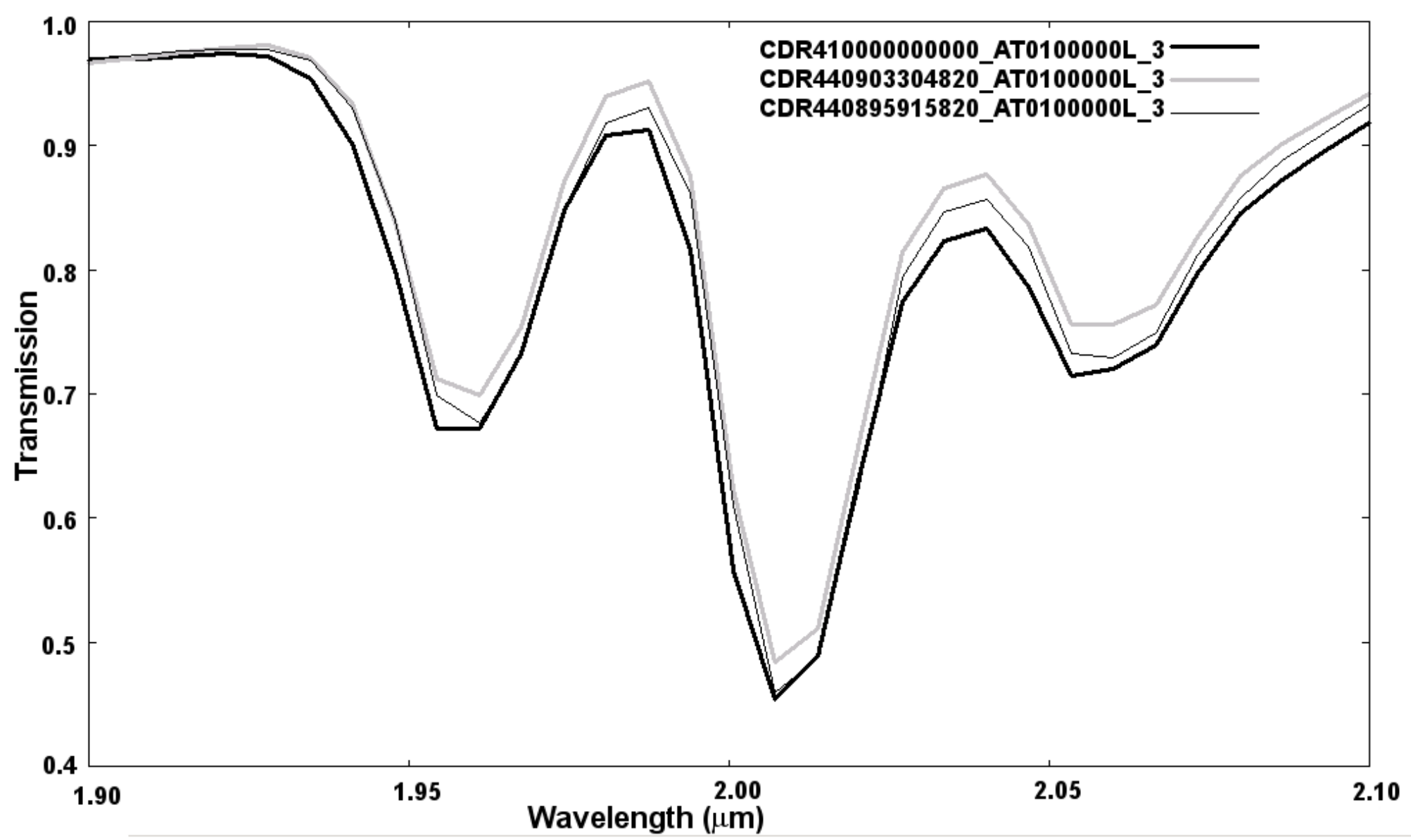

**Figure 4)** The 1.9-2.1 μm gas-band portion of volcano-scan spectra measured at three different times during the CRISM mission. There are 16 different volcano-scan spectra measured in total, but only three are shown here. Slight shifts in the wavelengths of the CRISM instrument (at the level of <0.001 μm) result in a slight modulation of the shape of these spectra. The transmission spectra were acquired at solar longitudes of 253° (Earth year 2007), 63° (Earth year 2008), and 30° (Earth year 2008), respectively, for the curves CDR410000000000, CDR440904404820 and CDR440895915820. Some of the temporal changes to the transmission spectra can be due to the effect of ice and/or dust aerosols, but for the transmission spectra shown here and for this wavelength range, the effect of the aerosols is reduced. The set of 16 such volcano-scan spectra constitute a database that is used for different parts of the mission in order to correct for the slight shifts in wavelength in the CRISM instrument.

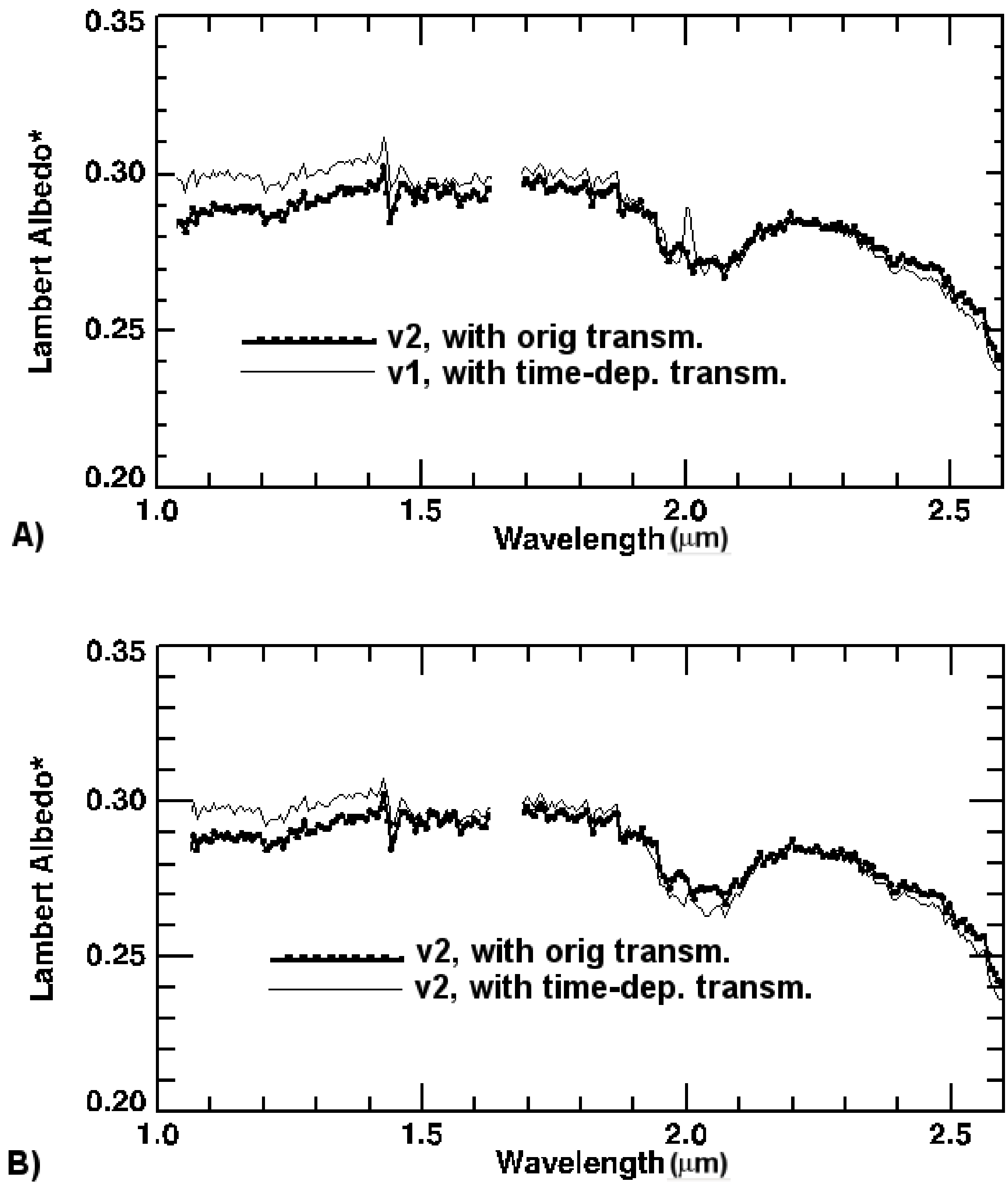

**Figure 5)** In panel A, the v1 algorithm with the new time-dependent volcano-scan transmission curve is compared to the v2 algorithm with the original volcano-scan transmission curve. In panel B, the v2 algorithm with the new time-dependent volcano-scan transmission curve is compared to the v2 algorithm with the original volcano-scan transmission curve. This is for the same spectrum shown in Fig. 1.

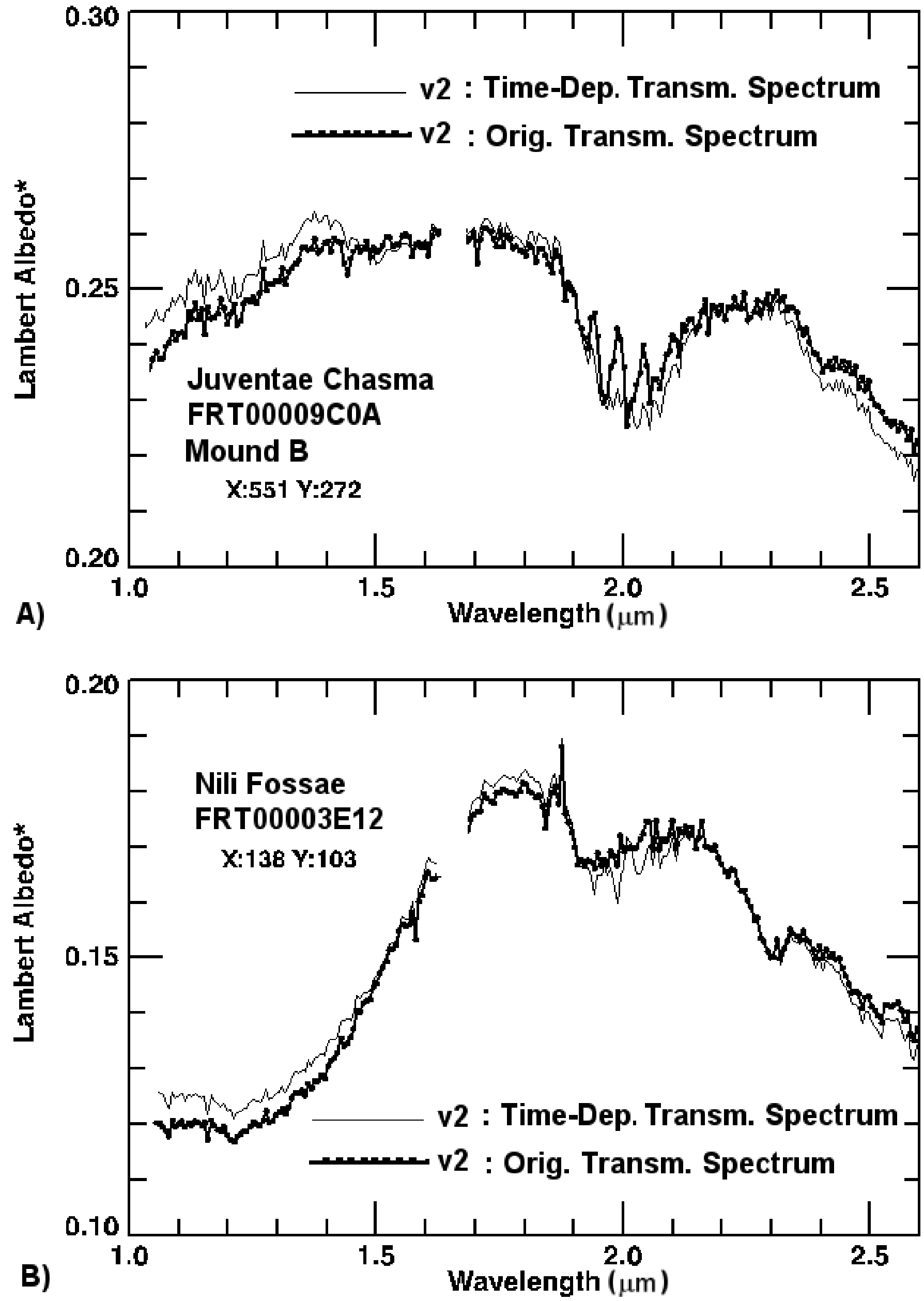

**Figure 6)** Comparison of corrected spectra using the v2 volcano-scan algorithm, with either the original transmission spectrum or the transmission spectra chosen in a time-dependent manner, for portions of two different CRISM spectra, for (panel A) a hyperspectral observation of a sulfate-rich area of Juventae Chasma, and (panel B) a hyperspectral observation of a phyllosilicate-rich area of Nili Fossae. The locations of these spectra are the same as in Fig. 2A and 2C. Note that when using the time-dependent transmission spectrum, the ripple near 2.0 μm in panel A is reduced by an order of magnitude, whereas the ripple near 2.0 μm in panel B is slightly higher than with the original transmission spectrum. At other wavelengths, differences in the spectra are largely caused by differences in the aerosol content during the measurement of the different volcano-scan transmission spectra.